# On the Validity of the Parabolic Effective-Mass Approximation for the Current-Voltage Calculation of Silicon Nanowire Transistors

Jing Wang, Anisur Rahman, Avik Ghosh, Gerhard Klimeck and Mark Lundstrom
School of Electrical and Computer Engineering, Purdue University, West Lafayette,
IN47907 (email: jingw@purdue.edu)

**Abstract**

This paper examines the validity of the widely-used parabolic effective-mass approximation for computing the current-voltage (*I-V*) characteristics of silicon nanowire transistors (SNWTs). The energy dispersion relations for unrelaxed Si nanowires are first computed by using an $sp^3d^5s^*$ tight-binding model. A semi-numerical ballistic FET model is then adopted to evaluate the *I-V* characteristics of the (n-type) SNWTs based on both a tight-binding dispersion relation and parabolic energy bands. In comparison with the tight-binding approach, the parabolic effective-mass model with bulk effective-masses significantly overestimates SNWT threshold voltages when the wire width is <3nm, and ON-currents when the wire width is <5nm. By introducing two analytical equations with two tuning parameters, however, the effective-mass approximation can well reproduce the tight-binding *I-V* results even at a ~1.36nm wire with.

# I. INTRODUCTION

As MOSFET gate lengths enter the sub-50nm regime, short channel effects become increasingly severe [1]. To further scale down MOSFETs, device structures with new gate configurations are preferred to provide better electrostatic control than planar structures. For this reason, silicon nanowire transistors (SNWTs), which allow multi-gate or gate-all-around structures, are being extensively explored by different experimental groups [2-5]. Rapid experimental progress in SNWTs has shown their potential applications in future electronics.

To understand the device physics of SNWTs and to assess their performance limits, simulation work is important. Recently, three-dimensional quantum mechanical simulations of (n-type) SNWTs (or FinFETs) have been accomplished based on the parabolic effective-mass approximation [6-8]. Due to the two-dimensional quantum confinement, however, the bulk crystal symmetry is not preserved in Si nanowires. For this reason, quantitative results obtained from the parabolic effective-mass approximation are expected to suffer errors when the nanowire diameter is small. In this work, we explore the validity of the parabolic effective-mass approximation for the current-voltage (*I-V*) calculation of silicon nanowire transistors. We first compute the energy dispersion (*E-k*) relations of Si nanowires by a semi-empirical $sp^3d^5s^*$ nearest-neighbor tight-binding approach [9-12]. The *I-V* characteristics of n-type SNWTs are then evaluated by a semi-numerical ballistic FET model [13-15] using both the tight-binding *E-k* relations and parabolic energy bands. By comparing the results for the two types of *E-k* relations, the validity of the parabolic effective-mass approximation is examined.

This paper is divided into the following sections. Sec. II describes the $sp^3d^5s^*$ tight-binding approach and illustrates the calculated atomistic nanowire bandstructures. In Sec. III, we first examine the validity of the parabolic effective-mass approximation for n-type SNWT simulations and then propose a tuning procedure to modify the effective-mass approach for a better agreement with the tight-binding calculation. Sec. IV summarizes key findings of this work.

## II. TIGHT-BINDING CALCULATION OF BANDSTRUCTRES

Figure 1 shows an example of the simulated nanowire structures in this work. The transport orientation of the wire is along the [100] direction (see Fig. 1 (a)), the shape of the cross section is square, and the faces of the square are all along the equivalent <100> axes (see Fig. 1 (c)). Fig. 1 (b) illustrates a unit cell of the nanowire crystal, which consists of four atomic layers along the $x$ (transport) direction and has a length of $a_0$=5.43Å. It should be noted that although Fig. 1 is only for a nanowire with a wire width $D$=1.36nm, nanowires with various wire widths (from 1.36nm to 6.79nm) are explored in this work.

According to the tight-binding approach adopted in this work, 20 orbitals, consisting of an $sp^3d^5s^*$ basis with spin-orbit coupling, are used to represent each atom in the nanowire Hamiltonian. The orbital-coupling parameters we use are from [9], which have been optimized by Boykin *et al.* to accurately reproduce the band gap and effective-masses of bulk Si. (It should be mentioned that bulk bond lengths are assumed in this work. In real nanowires, the crystal structures will relax to obtain a minimum energy [16]. We expect that the general results of this study will also apply to relaxed structures while some quantitative differences may appear.) At the Si surfaces, a hard wall boundary condition for the wavefunction is applied and the dangling bonds at these surfaces are passivated using a hydrogen-like termination model of the $sp^3$ hybridized interface atoms. As demonstrated in [17], this technique successfully removes all the surface states from the semiconductor band gap.

Figure 2 shows the *E-k* relations (left column) and the corresponding density-of-states (right column) for the simulated Si nanowires with wire widths (a) $D$=1.36nm and (b) $D$=5.15nm. It is clear that the six equivalent Δ valleys in the bulk Si conduction band split up into two groups due to quantum confinement. Four unprimed valleys, [010], [0$\bar{1}$0], [001], and [00$\bar{1}$], are projected to the Γ point ($k_x = 0$) in the one-dimensional wire Brillouin zone ($-\pi/a_0 \leq k_x \leq \pi/a_0$) to form the conduction band edge. Two primed valleys (i.e., [100] and [$\bar{1}$00]), located at $k_x = \pm 0.815 \cdot 2\pi$ in the bulk Brillouin zone, are zone-folded to $k_x = \pm 0.37\pi$ in the wire Brillouin zone to form the off-Γ states. A similar observation has been reported in [10] and [11] for square Si nanowires with a [100] transport direction and four confinement directions along the equivalent <110> axes. In the density-of-states (DOS) *vs.* energy plots (right column), peaks corresponding to each energy minimum (maximum) in the wire conduction (valence) band are clearly observed.

As in a Si quantum well, the degeneracy of the 4-fold Γ valleys in a [100] oriented square wire can be lifted by the interaction between the four equivalent valleys, which is so called "band splitting" [18, 19]. It is clearly seen in Fig. 2 that the band splitting is more significant in the thinner wire ($D$=1.36nm) than in the thicker wire ($D$=5.15nm). Fig. 3 plots the wire width ($D$) dependence of the splitting energy, defined as the difference between the highest and the lowest energy (at the Γ point) of the four split conduction bands. The splitting energy is seen to fluctuate and the envelope decreases with the wire width according to $D^{-3}$, analogous with the band splitting observed in Si quantum wells [18, 19].

### III. VALIDITY OF THE PARABOLIC EFFECTIVE-MASS APRROACH

In this section, we adopt a semi-numerical ballistic FET model to calculate the *I-V* characteristics of n-type SNWTs based on both tight-binding *E-k* relations and parabolic energy bands. The main features of the ballistic FET model are illustrated in Fig. 4. Three capacitors, $C_G$, $C_S$, and $C_D$ are employed to describe the electrostatic couplings between the top of the barrier and the gate, source and drain terminals, respectively. The potential at the top of the barrier is obtained as

$$U_{scf} = \left(\frac{C_G}{C_G+C_D+C_S}\right)V_G + \left(\frac{C_D}{C_G+C_D+C_S}\right)V_D + \left(\frac{C_S}{C_G+C_D+C_S}\right)V_S + \frac{Q_{Top}}{(C_G+C_D+C_S)}, \quad (1)$$

where $V_G$, $V_S$, and $V_D$ are the applied biases at the gate, the source and the drain, respectively, and $Q_{Top}$ is the mobile charge at the top of the barrier, which is determined by $U_{scf}$, the source and drain Fermi levels ($E_{FS}$ and $E_{FD}$) and the $E$-$k$ relation for the channel material. To be specific, the group velocity of each state is calculated from the tabulated $E$-$k$ data of the nanowire, and the carrier density is then evaluated by assuming that the states with a positive (negative) group velocity are in equilibrium with the source (drain) reservoir. After self-consistency between $U_{scf}$ and $Q_{Top}$ is achieved, the drain current is readily obtained from the known populations of all the states in the energy bands of the wire. In previous work, this model was used to evaluate the $I$-$V$ characteristics of ballistic Si MOSFETs [13] and HEMTs [14] with parabolic energy bands and Ge MOSFETs with numerical $E$-$k$ relations [15]. A detailed description of the model can be found in [13] and the Matlab® scripts of this model are available [20].

Figure 5 plots the $I_{DS}$ vs. $V_{GS}$ curves for a square SNWT with $D$=1.36nm in both (a) a semi-logarithmic scale and (b) a linear scale. The dashed lines are for the results based on the tight-binding $E$-$k$ relations while the solid lines are for the parabolic effective-mass (pEM) results. In the parabolic effective-mass approach, all six conduction-band valleys in bulk Si are considered, and the effective-masses used in the calculation ($m_l = 0.891 m_e$ and $m_t = 0.201 m_e$) are extracted from the bulk $E$-$k$ relation evaluated by our tight-binding approach with the parameters obtained from [9]. If we define a threshold voltage, $V_T$, as

$$I_{DS}(V_{GS}@V_T, V_{DS}=0.4\text{V})=300\text{nA}, \quad (2)$$

and an ON-current of SNWTs as

$$I_{ON}=I_{DS}(V_{GS}-V_T=0.3\text{V}, V_{DS}=0.4\text{V}), \qquad (3)$$

we find that pEM significantly overestimates the threshold voltage by $V_T^{pEM} - V_T^{TB} = 0.28V$ and the ON-current by $(I_{ON}^{pEM} - I_{ON}^{TB})/I_{ON}^{TB} = 42\%$ as compared with the tight-binding results. Fig. 6 compares pEM (solid) vs. tight-binding (circles) for the *I-V* calculation of a thicker SNWT with *D=6.79*nm. It is clear that pEM provides nearly identical *I-V* characteristics as tight-binding except for a small overestimation of ON-current by ~5%. The solid lines with circles in Fig. 7 show the wire width (*D*) dependence of the errors, $V_T^{EM} - V_T^{TB}$ in (a) and $(I_{ON}^{EM} - I_{ON}^{TB})/I_{ON}^{TB}$ in (b), associated with pEM. It is clear that pEM starts to overestimate threshold voltage by >0.03V when *D* scales below 3nm and ON-current by ≥10% when *D* is ≤5nm.

To understand the above observations, we plot the *D* dependence of the wire conduction band-edges, $E_C$, and the transport effective-mass, $m_x^*$, at the Γ point in the wire conduction band (Fig. 8). As we can see in Fig. 8 (a), when *D*>4nm, the $E_C$ obtained from the tight-binding calculations (solid with circles) is well reproduced by pEM (dashed). (In pEM, the wire conduction band-edge is determined by the lowest subband level of the four unprimed valleys). At smaller wire widths, however, pEM overestimates $E_C$ due to the nonparabolicity of the bulk Si bands [21, 22]. This overestimation of $E_C$ by pEM directly leads to the overvalued threshold voltages of the simulated SNWTs. The solid line with circles in Fig. 8 (b) shows an increasing $m_x^*$ (extracted from the tight-binding *E-k* relations) with a decreasing *D*, which is also a result of the nonparabolicity of the bulk Si *E-k* relations. When *D*<3nm, $m_x^*$ extracted from tight-binding is >50% larger than the corresponding bulk value used in pEM. Since the electron thermal velocity is inversely proportional to the square root of the transport effective-mass, the parabolic effective-mass calculations, which adopt a smaller $m_x^*$ than the tight-binding approach, overestimate the carrier injection velocity and consequently the SNWT ON-currents. In short, the nonparabolicity of the bulk Si bands plays an important role when quantum confinement is strong (small *D*). The use of parabolic energy bands overestimates the wire conduction band-edge and underestimates the transport effective-mass, and consequently provides a higher SNWT threshold voltage and ON-current as compared with the tight-binding approach.

Although we have shown that the *parabolic* effective-mass approach does not perform well at small wire widths, it is still interesting to know whether it is possible to modify the effective-mass approach to obtain a better agreement with the tight-binding calculation, since the effective-mass approximation significantly reduces computation time as compared to atomistic treatments. To do this, we first define a quantum confinement energy as the difference between the wire conduction band-edge, $E_C$ and that for bulk Si ($E_C^{bulk} = 1.13 eV$). Fig. 9 (a) shows the quantum confinement energy computed by pEM ($E_{QC}^{pEM}$) *vs.* that obtained from the tight-binding calculation ($E_{QC}^{TB}$). It is evident that for small wire widths, the data points (circles) stand above the *y=x* curve, indicating that pEM overestimates the quantum confinement energy when it is large. Inspired by the expressions for the nonparabolicity of the bulk Si bands [21, 22], we propose the following quadratic equation to analytically describe the $E_{QC}^{pEM}$ *vs.* $E_{QC}^{TB}$ relation,

$$E_{QC}^{TB} \cdot \left(1 + \alpha \cdot E_{QC}^{TB}\right) = E_{QC}^{pEM}, \tag{4}$$

where $\alpha$ is treated as a fitting parameter and $\alpha = 0.27 eV^{-1}$ is used for the solid line in Fig. 9 (a) for the best agreement with the extracted data. Similarly, the $E_{QC}^{TB}$ dependence of the transport effective-mass $m_x^*$ at the Γ point (Fig. 9 (b)) can also be described by the following equation,

$$m_x^* = m_{bulk}^*\left(1 + \beta \cdot E_{QC}^{TB}\right), \tag{5}$$

where $m_{bulk}^* = m_t = 0.201 m_e$ is the transport effective-mass in the unprimed valleys in bulk Si and $\beta = 1.5 eV^{-1}$ is chosen to achieve the best match between the extracted data points (circles) and the analytical expression (solid).

After knowing Eqs. (4) and (5), the effective-mass approximation can be tuned for a better fit with tight-binding in the following steps.

Step 1) Calculate the quantum confinement energy, $E_{QC}^{pEM}$ by the parabolic effective-mass approach with the bulk effective-masses (i.e., $m_y^*$ and $m_z^*$).

Step 2) Solve Eq. (4) to obtain the updated quantum confinement energy, $E_{QC}^{new}$, as

$$E_{QC}^{new} = \frac{-1 + \sqrt{1 + 4\alpha \cdot E_{QC}^{pEM}}}{2\alpha}. \qquad (6)$$

Step 3) Evaluate the tuned transport effective-mass at the Γ point by Eq. (5),

$$m_x^* = m_{bulk}^* \left(1 + \beta \cdot E_{QC}^{new}\right). \qquad (7)$$

Step 4) Use the computed $E_{QC}^{new}$ and $m_x^*$ for the *I-V* calculation of SNWTs.

It should be noted that the above tuning process is only necessary for the four unprimed valleys because i) at large wire widths, the quantum confinement energy is small and nonparabolicity is insignificant in both unprimed and primed valleys, so the parabolic effective-mass approach performs well, and ii) at small wire widths, the two primed valleys are well separated from the unprimed ones due to stronger quantum confinement (smaller effective-masses in the *y* and *z* directions) in these primed valleys, so the electron density and current contributed by the primed valleys are negligible.

The dashed lines with diamonds in Fig. 7 show the wire width (*D*) dependence of the errors, $V_T^{EM} - V_T^{TB}$ in (a) and $\left(I_{ON}^{EM} - I_{ON}^{TB}\right)/I_{ON}^{TB}$ in (b), associated with the *tuned* effective-mass approximation. For wire widths ranging from 1.36nm to 6.79nm, the tuned effective-mass approach provides an excellent match with the tight-binding calculation – less than 10mV error for $V_T$ and less than 5% error for $I_{ON}$. So far, we have shown that the effective-mass approximation can be modified by introducing two *D*-independent parameters, $\alpha$ and $\beta$, to accurately reproduce the *I-V* results computed by tight-binding. It must be mentioned that the *values* of $\alpha$ and $\beta$ used in this work were obtained for SNWTs with one particular channel orientation (i.e., [100]) and one specific cross-sectional shape (i.e., square with all faces along the equivalent <100> axes). The important point is that for *I-V* calculation it is possible to simply tune the effective-mass approach to fit the tight-binding model. We expect that this conclusion may apply to other SNWTs with different transport directions and cross sections while the *values* of the tuning parameters ($\alpha$ and $\beta$) are subject to change.

## IV. CONCLUSIONS

By using an $sp^3d^5s^*$ tight-binding approach as a benchmark, we examined the validity of the parabolic effective-mass approximation for the current-voltage calculation of n-type silicon nanowire transistors. It was found that the simple parabolic effective-mass approach with bulk effective-masses significantly overestimates SNWT threshold voltages when the wire width (*D*) is <3nm, and ON-currents when *D*<5nm. However, by introducing two analytical equations with two tuning parameters, the effective-mass approximation can well reproduce the tight-binding *I-V* results over a wide range of wire widths – even at *D*=1.36nm. In conclusion, bandstructure effects begin to manifest themselves in silicon nanowires with small diameters, but with a simple tuning procedure, the parabolic effective-mass approximation may still be used to assess of the performance limits of silicon nanowire transistors.

ACKNOWLEDGEMENTS

This work is supported by the Semiconductor Research Corporation (SRC), and the NSF Network for Computational Nanotechnology (NCN). The authors thank Prof. Supriyo Datta at Purdue University for fruitful discussions.

**FIGURE CAPTIONS:**

Fig. 1   (a) The atomic structure of a square nanowire ($D$=1.36nm) with a [100] transport direction. (b) A unit cell of the square nanowire illustrated in (a). (c) The schematic diagram of the cross section of the square nanowire. $D$ demotes the edge length of the square cross section and the four faces of the square are all along the equivalent <100> axes.

Fig. 2   The tight-binding $E$-$k$ relations and the corresponding density-of-states (DOS) for the simulated Si nanowire structures with (a) $D$=1.36nm and (b) $D$=5.15nm. The conduction band for the thinner wire ($D$=1.36nm) displays significant band splitting at the $\Gamma$ point.

Fig. 3   The splitting energy (at the $\Gamma$ point) *vs.* wire width ($D$) for the simulated Si nanowires. The closed circles are for the wires with an odd number of atomic layers while the open circles are for the ones with an even number of atomic layers. The splitting energy fluctuates with $D$ and the envelope decreases according to $\sim D^{-3}$.

Fig. 4   Illustration of the essential aspects of the semi-numerical ballistic FET model.

Fig. 5   The $I_{DS}$ *vs.* $V_{GS}$ curves for a square SNWT with $D$=1.36nm in both (a) a semi-logarithmic scale and (b) a linear scale. The oxide thickness is 1nm, the temperature is 300K, and the drain bias is 0.4V. The dashed lines are for the results based on the tight-binding (TB) $E$-$k$ relations while the solid lines for the parabolic effective-mass (pEM) results.

Fig. 6   The $I_{DS}$ *vs.* $V_{GS}$ curves for a square SNWT with $D$=6.79nm in both a semi-logarithmic scale (left) and a linear scale (right). The oxide thickness is 1nm, the temperature is 300K, and the drain bias is 0.4V. The circles are for the results based on the tight-binding (TB) $E$-$k$ relations while the solid lines for the parabolic effective-mass (pEM) results.

Fig. 7   The wire width dependence ($D$) of the errors, $V_T^{EM} - V_T^{TB}$ in (a) and $\left(I_{ON}^{EM} - I_{ON}^{TB}\right)/I_{ON}^{TB}$ in (b), associated with the effective-mass approximations. The solid lines with circles are for the parabolic effective-mass (pEM) approximation while the dashed lines with diamonds are for the tuned effective-mass (tEM) approach.

Fig. 8 (a) The conduction band edges, $E_C$, for the simulated wires with different wire widths. The solid line with circles is for the values obtained from the tight-binding (TB) $E$-$k$ relations and the dashed line is for the parabolic effective-mass (pEM) results. (b) The wire width ($D$) dependence of the transport effective-mass, $m_x^*$, at the Γ point in the wire conduction band (extracted from the tight-binding energy bands by $m_x^* = \hbar^2 /(\partial^2 E / \partial^2 k_x)$, where $\hbar$ is the Plank constant). For comparison, the bulk value of $m_x^*$ for the unprimed valleys (used in pEM) is shown by the dash-dot line.

Fig. 9 (a) The quantum confinement energy computed by parabolic effective-mass ($E_{QC}^{pEM}$) vs. that obtained from the tight-binding calculation ($E_{QC}^{TB}$). (b) The ratio of the transport effective-mass, $m_x^*$ to the bulk value, $m_x^{bulk}$ vs. the quantum confinement energy calculated by the tight-binding approach ($E_{QC}^{TB}$). In both plots, the circles are for the data points extracted from the tight-binding and parabolic $E$-$k$ relations, and the corresponding wire width for each point from left to right is D=6.79nm, 5.15nm, 3.53nm, 1.90nm, and 1.36nm, respectively. The solid lines are for the analytical fit based on Eqs. (4) and (5).

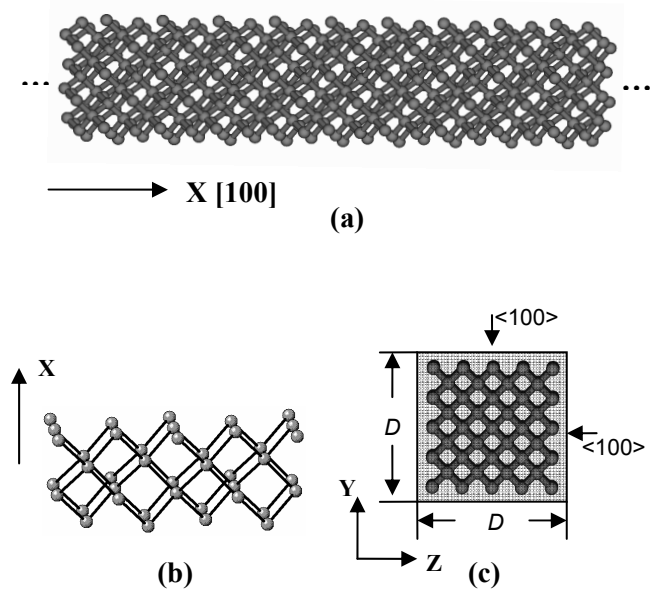

Fig. 1 (a) The atomic structure of a square nanowire (*D*=1.36nm) with a [100] transport direction. (b) A unit cell of the square nanowire illustrated in (a). (c) The schematic diagram of the cross section of the square nanowire. *D* demotes the edge length of the square cross section and the four faces of the square are all along the equivalent <100> axes.

Fig. 2 Wang, Rahman, Ghosh, Klimeck and Lundstrom

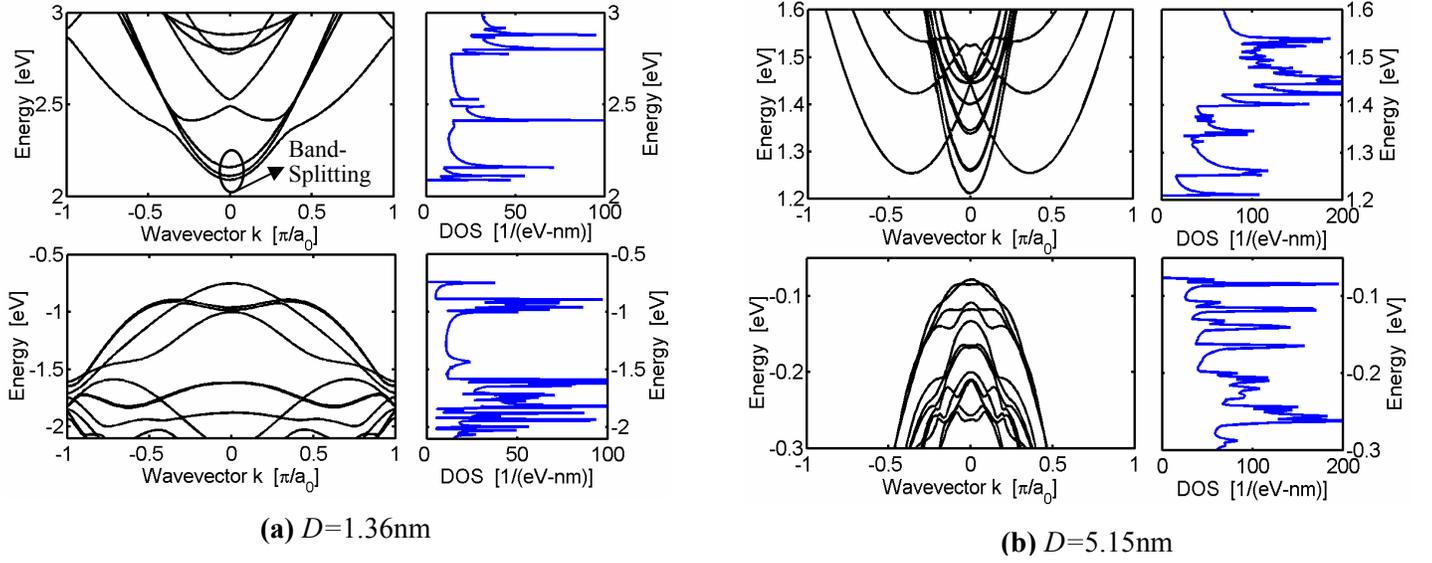

**(a)** $D$=1.36nm

**(b)** $D$=5.15nm

Fig. 2 The tight-binding $E$-$k$ relations and the corresponding density-of-states (DOS) for the simulated Si nanowire structures with (a) $D$=1.36nm and (b) $D$=5.15nm. The conduction band for the thinner wire ($D$=1.36nm) displays significant band splitting at the $\Gamma$ point.

Fig. 3 Wang, Rahman, Ghosh, Klimeck and Lundstrom

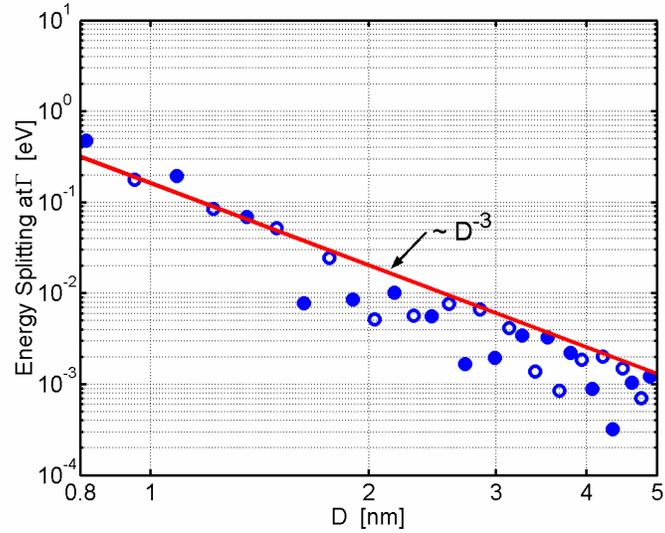

Fig. 3 The splitting energy (at the Γ point) *vs.* wire width (*D*) for the simulated Si nanowires. The closed circles are for the wires with an odd number of atomic layers while the open circles are for the ones with an even number of atomic layers. The splitting energy fluctuates with *D* and the envelope decreases according to ~$D^{-3}$.

Fig. 4 Wang, Rahman, Ghosh, Klimeck and Lundstrom

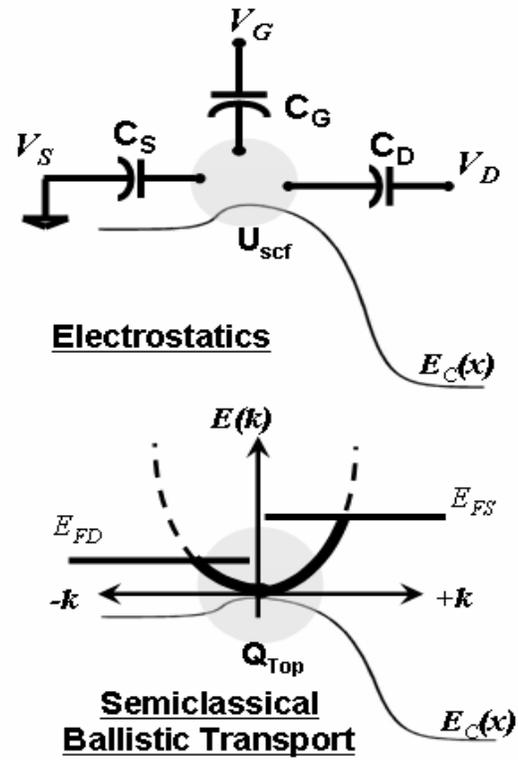

Fig. 4 Illustration of the essential aspects of the semi-numerical ballistic FET model.

Fig. 5 Wang, Rahman, Ghosh, Klimeck and Lundstrom

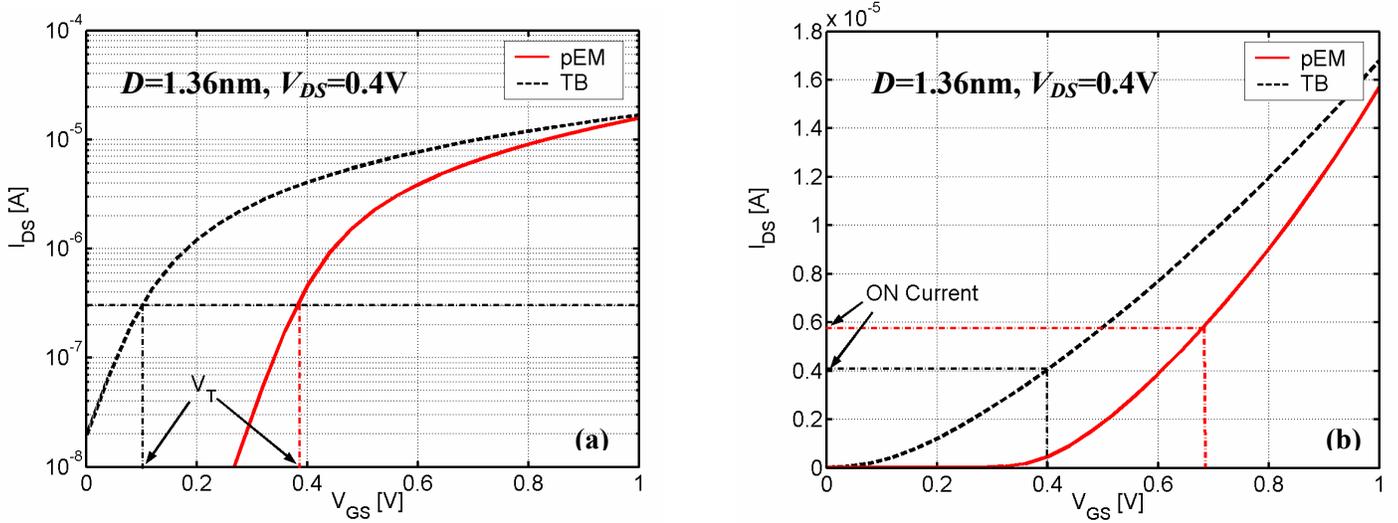

Fig. 5 The $I_{DS}$ vs. $V_{GS}$ curves for a square SNWT with $D$=1.36nm in both (a) a semi-logarithmic scale and (b) a linear scale. The oxide thickness is 1nm, the temperature is 300K, and the drain bias is 0.4V. The dashed lines are for the results based on the tight-binding (TB) $E$-$k$ relations while the solid lines for the parabolic effective-mass (pEM) results.

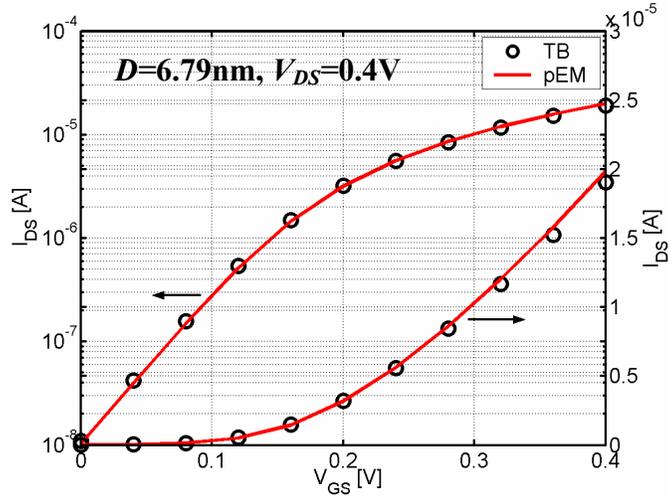

Fig. 6 The $I_{DS}$ vs. $V_{GS}$ curves for a square SNWT with $D$=6.79nm in both a semi-logarithmic scale (left) and a linear scale (right). The oxide thickness is 1nm, the temperature is 300K, and the drain bias is 0.4V. The circles are for the results based on the tight-binding (TB) $E$-$k$ relations while the solid lines for the parabolic effective-mass (pEM) results.

Fig. 7 Wang, Rahman, Ghosh, Klimeck and Lundstrom

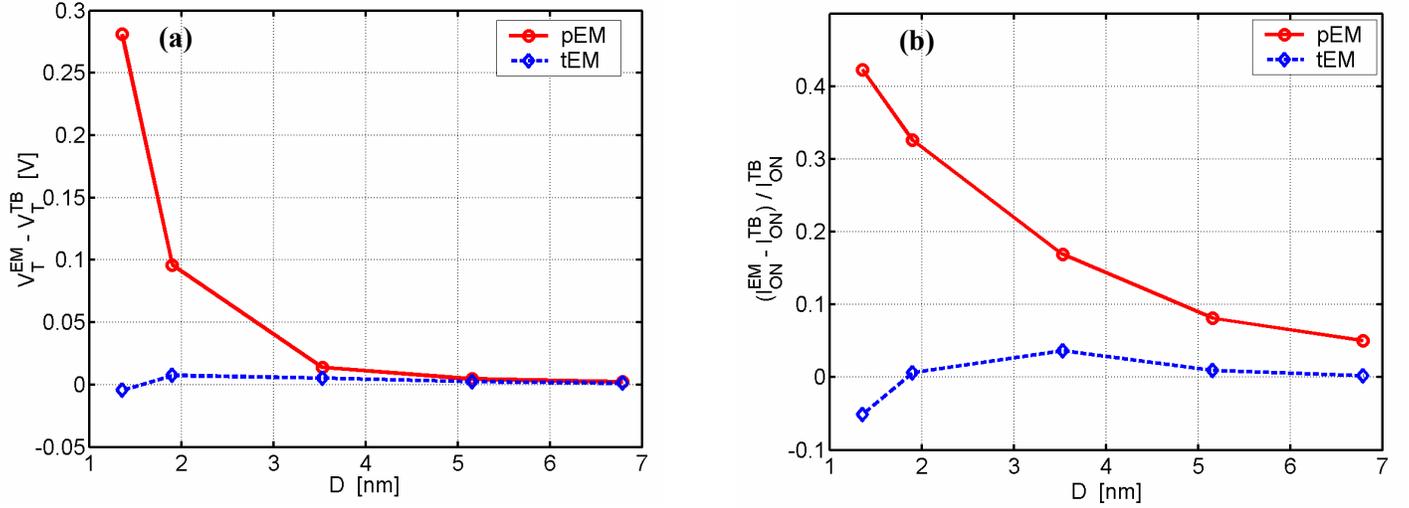

Fig. 7 The wire width dependence ($D$) of the errors, $V_T^{EM} - V_T^{TB}$ in (a) and $\left(I_{ON}^{EM} - I_{ON}^{TB}\right)/I_{ON}^{TB}$ in (b), associated with the effective-mass approximations. The solid lines with circles are for the parabolic effective-mass (pEM) approximation while the dashed lines with diamonds are for the tuned effective-mass (tEM) approach.

Fig. 8 Wang, Rahman, Ghosh, Klimeck and Lundstrom

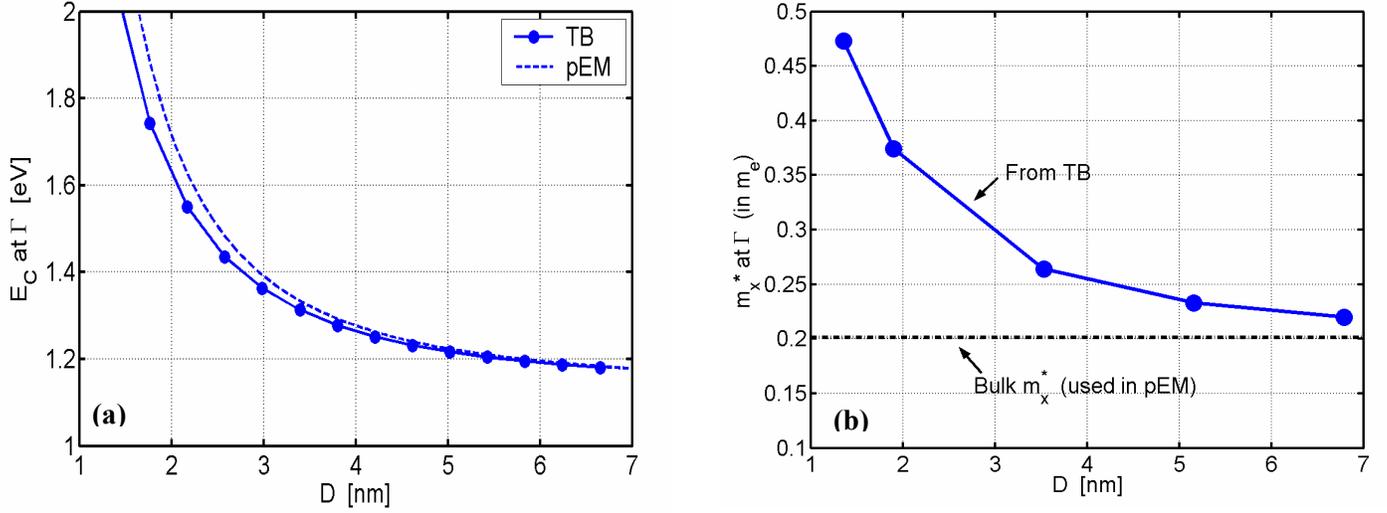

Fig. 8 (a) The conduction band edges, $E_C$, for the simulated wires with different wire widths. The solid line with circles is for the values obtained from the tight-binding (TB) $E$-$k$ relations and the dashed line is for the parabolic effective-mass (pEM) results. (b) The wire width ($D$) dependence of the transport effective-mass, $m_x^*$, at the Γ point in the wire conduction band (extracted from the tight-binding energy bands by $m_x^* = \hbar^2 / (\partial^2 E / \partial^2 k_x)$, where $\hbar$ is the Plank constant). For comparison, the bulk value of $m_x^*$ for the unprimed valleys (used in pEM) is shown by the dash-dot line.

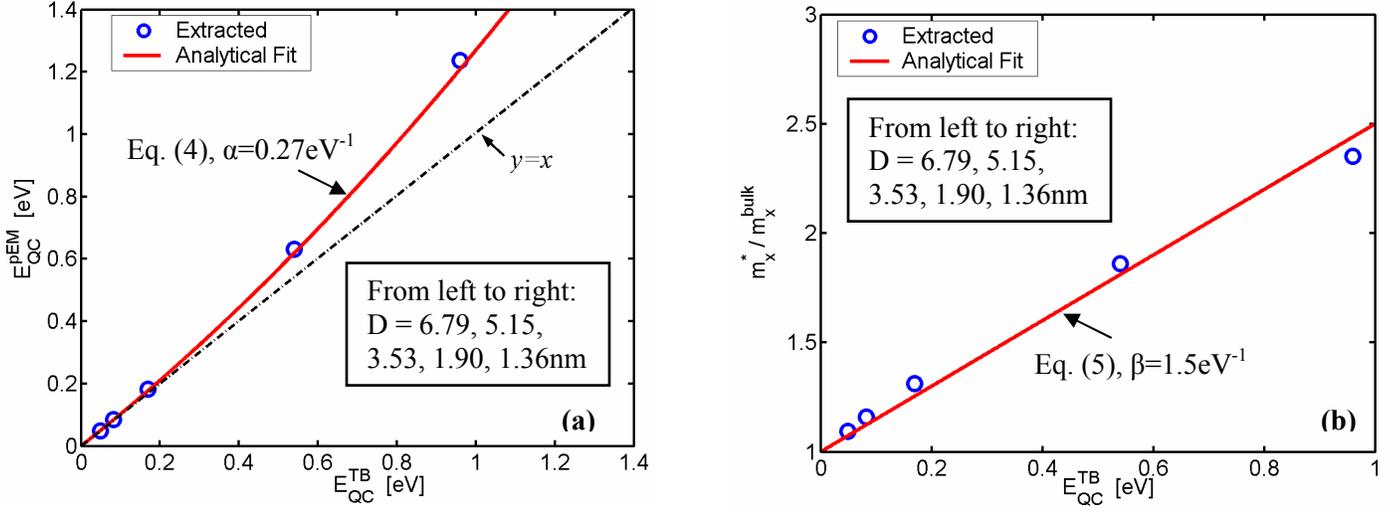

Fig. 9 (a) The quantum confinement energy computed by parabolic effective-mass ($E_{QC}^{pEM}$) vs. that obtained from the tight-binding calculation ($E_{QC}^{TB}$). (b) The ratio of the transport effective-mass, $m_x^*$ to the bulk value, $m_x^{bulk}$ vs. the quantum confinement energy calculated by the tight-binding approach ($E_{QC}^{TB}$). In both plots, the circles are for the data points extracted from the tight-binding and parabolic E-k relations, and the corresponding wire width for each point from left to right is D=6.79nm, 5.15nm, 3.53nm, 1.90nm, and 1.36nm, respectively. The solid lines are for the analytical fit based on Eqs. (4) and (5).